\documentclass[%
 reprint,
 amsmath,amssymb,
 aps,
pra,
]{revtex4-2}

\usepackage{bm,physics,amsmath,amssymb,color}
\usepackage[utf8]{inputenc}
\usepackage[top=0.6in, bottom=1in, left=0.8in, right=0.8in]{geometry}
\usepackage[normalem]{ulem}
\usepackage{graphicx,layouts}

\begin{document}

\title{Massive Dirac-Pauli physics in lead-halide perovskites}%

\author{Abhishek Shiva Kumar$^1$}

\author{Mikhail Maslov$^1$}

\author{Mikhail Lemeshko$^1$}

\author{Artem G. Volosniev$^{1,2}$}
\email{artem@phys.au.dk}

\author{Zhanybek Alpichshev$^1$}
\email{alpishev@ist.ac.at}

\affiliation{$^1$Institute of Science and Technology Austria (ISTA), Am Campus 1, 3400 Klosterneuburg, Austria}
\affiliation{$^2$Department of Physics and Astronomy, Aarhus University, Ny Munkegade 120, DK-8000 Aarhus C, Denmark
}%

\date{\today}

\begin{abstract}
 In standard quantum electrodynamics (QED), the so-called non-minimal (Pauli) coupling is suppressed for elementary particles and has no physical implications. Here, we show that the Pauli term naturally appears in a known family of Dirac materials --
 the lead-halide perovskites, suggesting a novel playground for the study of analogue QED effects. We outline 
measurable
manifestations of the Pauli term in the phenomena pertaining to (i) relativistic corrections to bound states (ii) the Klein paradox, {\color{black} and (iii) spin effects in scattering}. In particular, we demonstrate that (a) the
binding energy of an electron in the vicinity of a positively charged defect is noticeably decreased due to the polarizability of lead ions and the appearance of a Darwin-like term, {\color{black} (b) strong spin-orbit coupling due to the Pauli term affects the exciton states, and (c) scattering of an electron off an energy barrier with broken mirror symmetry produces spin polarization in the outgoing current}.
 Our study adds to understanding of quantum phenomena in lead-halide perovskites, and
 paves the way for tabletop simulations of analogue Dirac-Pauli equations.
\end{abstract}

\maketitle

 The band structure of certain condensed matter systems (Dirac materials) resembles the energy spectrum of the single-body relativistic problem~\cite{Wehling2014}.
Arguably, the best known system of this type is graphene whose massless Dirac-like band structure leads to a number of remarkable physical phenomena that were previously discussed  mainly in the context of relativistic high-energy physics~\cite{CastroNeto2009}. Topological insulators and Weyl semimetals are among the other celebrated examples~\cite{Hasan2010,Qi2011,Armitage2018}. These materials have become the tabletop experimental platforms for visualizing the extreme relativistic phenomena that obey conventional particle theories, such as QED. The natural question that may arise in this context is whether the analog Dirac materials can help to study conceivable high-energy effects that fall beyond the Standard Model of elementary particles.

The coupling  of Dirac materials to electromagnetic fields is typically given via a minimal substitution: $p^\nu\rightarrow p^\nu-q A^\nu$, where $p^\nu$ and $A^\nu=(\phi, \bm{A})$ are the usual four-momentum and  electromagnetic four-potential, respectively~\cite{Peskin1995}. 
However, Lorentz and gauge invariance allow in general for additional `non-minimal' couplings to electromagnetic fields (see, e.g., Ref.~\cite{Foldy1952}). The first such extension of the Dirac equation was suggested by Pauli~\cite{Pauli1941} who introduced what is now known as the Dirac-Pauli equation~\cite{BagrovGitman+2014}
\begin{equation}
   \left (\gamma^\nu (p_\nu-q A_\nu)- \frac{i \mu'}{4} [\gamma_{\mu},\gamma_{\nu}] F^{\mu \nu}-m\right)\psi = 0.
    \label{DP equation}
\end{equation}
 Here, $\gamma^\nu$ are the Dirac matrices, $F^{\mu \nu}$ is the electromagnetic field tensor, $m$ is the rest mass of the Dirac particle and $\mu'$ is its anomalous moment. The experimental data on magnetic moments of elementary particles suggest that $\mu'$ is  negligibly small~\cite{Weinberg1996}. It can even be assumed that $\mu'=0$ at the current state of our understanding of quantum electrodynamics~\footnote{Recently, it was proposed that the Pauli term can be important in the context of the Landau  pole~\cite{Djukanovic2018,Gies2020}. This however does not change the fact that the value of $\mu'$ must be negligibly small.}. 
 Therefore, the Pauli term is typically neglected with rare exceptions of effective quantum field theories that aim to include polarizabilities of composite particles such as neutrons and protons~\cite{Powell1949,PhysRevLett.3.105,Bawin1997} in a phenomenological way.
 
 In this paper, we show that the Pauli term naturally appears in the low-energy description of lead-halide perovskites (LHPs), making it (to the best of our knowledge) the first known Dirac-Pauli semiconductor. 
  Lead-halide perovskites are a family of advanced energy materials~\cite{Grtzel2014} with exceptional optoelectronic properties~\cite{Zhao2016,Kovalenko2017}. Charge carriers in LHPs can be described by the massive Dirac equation~\cite{Jin2012,Becker2018} with a peculiarity that their coupling to electromagnetic fields requires an explicit inclusion of the high polarizability of Pb$^{2+}$ ions~\cite{PhysRevB.107.125201,PhysRevLett.130.106901}.

\vspace{1em}

\noindent {\large \bf Results}

\noindent Here, we first illustrate that the non-standard matter-field interaction in LHPs is actually the Pauli coupling with the (effective) atomic dipole polarizability in place of the anomalous moment, $\mu'$. Further, to facilitate future theoretical and experimental studies, we outline a few effects that the Dirac-Pauli physics can have on the properties of LHPs. {\color{black} We consider bound states of charge carriers to local defects, such as vacancies and interstitials, and also electron-hole binding. Finally, we discuss physics of scattering in presence of the Pauli term}.

\vspace{1em}

\noindent {\bf Pauli coupling in LHPs}

\noindent The Hilbert space for the low-energy description of LHPs is determined by the hybridization of orbitals of
lead and halide atoms~\cite{Umebayashi2003}. In the vicinity of the chemical potential, the states
have predominantly $s$- (valence band) and $p$-type (conduction band) characters~\cite{Jin2012,Becker2018}. Strong spin-orbit coupling splits
the $p$-type states. Therefore, both the top of the valence band and the bottom of the conduction band have $J = 1/2$, and low-energy optoelectronic properties of LHPs are shaped by four states that can be conveniently written as $\{\Uparrow\uparrow,\Downarrow\uparrow,\Uparrow\downarrow,\Downarrow\downarrow\}$, where
\begin{equation}
\begin{split}
 |\!\!\Uparrow\uparrow\rangle = -\left(|p_z\rangle |\!\!\uparrow\rangle +\left( |p_x\rangle +i |p_y\rangle \right) |\!\!\downarrow\rangle\right)/\sqrt{3},\\
 |\!\!\Uparrow\downarrow\rangle = \left(|p_z\rangle |\!\!\downarrow\rangle -\left( |p_x\rangle -i |p_y\rangle \right) |\!\!\downarrow\rangle\right)/\sqrt{3},\\
 |\!\!\Downarrow\uparrow\rangle =|s\rangle|\!\!\uparrow\rangle, \quad \quad \quad |\!\!\Downarrow\downarrow\rangle =|s\rangle|\!\!\downarrow\rangle.
\end{split}
\label{eq:states}
\end{equation}
The right-hand-side of these expressions follows the standard notation~\cite{Kane1966,Chuang1995} for the spin structure of the state [$|\downarrow\rangle$, $|\uparrow\rangle$] and for its spatial component [$|s\rangle$, $|p_x\rangle$, $|p_y\rangle$, $|p_z\rangle$]. 
The Hermitian operators that act in the Hilbert space of Eq.~(\ref{eq:states}) can be readily written as the direct product $\tau_i \otimes \sigma_j$, where $\tau_i$ and $\sigma_j$ are the Pauli matrices that operate correspondingly in the orbital ($\{\Uparrow,\Downarrow\}$) and the quasi-spin ($\{\uparrow,\downarrow\}$) subspaces defined via the left-hand-side of the identities in  Eq.~(\ref{eq:states}). 

In the absence of external fields, the low-energy spectrum of LHPs is determined by diagonalizing the Hamiltonian ($\hbar=c=1$)~\cite{Jin2012}:
\begin{equation}
    H=  \Delta(\bm{k})\tau_3 \otimes \mathbb{I}+ 2at\ \tau_2 \otimes \bm{\sigma} \cdot \bm{k},
    \label{k.p hamiltonian}
\end{equation}
    where $\bm{\sigma}$ is a vector of Pauli matrices, so that $\bm{\sigma} \cdot \bm{k}=\sum_i \sigma_i k_i$. $\Delta(\bm{k}) =\frac{1}{2}\bigg(\Delta +  \frac{ t_{3} a^2 |\bm{k}|^2}{2}\bigg)$,
    $\bm{k}$ is the quasi-momentum counted from the high-symmetry $R$-point relevant for low-energy physics of LHPs, 
  $\Delta$ is the bandgap at the $R$-point, $a$ is the lattice constant, $t_3$ and $t$ are further $\bm{k}\cdot \bm{p}$ parameters of the Hamiltonian. They can be estimated by fitting to numerical calculations or experimental data. For example, for methylammonium lead bromide (MAPbBr$_3$) these parameters are found to be: $t\simeq 0.6$eV, $t_3\simeq0.9$eV, $a\simeq0.586$nm, $\Delta\simeq2.3$eV~\cite{Becker2018,PhysRevLett.130.106901}.

It was recently shown both experimentally and theoretically that in order to account for the full effect of the external electric field $\bm{E}$ on LHPs, one should employ not only the minimal coupling $\bm{\Tilde{k}}=\bm{k}-q\bm{A}$ (where $q$ is the charge of an electron) in Eq.~(\ref{k.p hamiltonian}) but also an additional term $\mu\ \tau_1 \otimes\bm\sigma \cdot \bm{E}$, dubbed `spin-electric' in Refs.~\cite{PhysRevB.107.125201,PhysRevLett.130.106901} because it couples the quasi-spin degree of freedom to an external electric field. The non-minimal coupling term accounts for the local effects of the electric field on the lattice, in particular, on the highly polarizable Pb$^{2+}$. Therefore, we interpret $\mu$ as coming from the effective atomic polarizability of  lead ions. For MAPbBr$_3$, the experiment demonstrates that the magnitude of this effect corresponds to $\mu\simeq-0.3|q|a$~\cite{PhysRevLett.130.106901}.

A close examination reveals that the operator $\mu\ \tau_1 \otimes\bm\sigma \cdot \bm{E}$ is nothing else but the electric part of the Pauli term in Eq.~(\ref{DP equation}). Indeed, the Hamiltonian in Eq.~(\ref{k.p hamiltonian}) amended with the spin-electric term can be written as:
\begin{equation}
    H= \Delta(\bm{\Tilde{k}}) \beta + 2at \bm{\alpha}\cdot\bm{\Tilde{k}}+i \mu \bm{\gamma} \cdot \bm{E} ,
    \label{dirac hamiltonian}
\end{equation}
where $\beta=\tau_3\otimes \mathbb{I}$, $\bm{\alpha}=\tau_2\otimes\bm{\sigma}$, and  $\bm{\gamma}=\beta  \bm{\alpha}$. As the  matrices $\gamma^{\nu}=\{\beta,\bm{\gamma}\}$ satisfy the Dirac algebra
$\{ \gamma^{\mu},\gamma^{\rho} \}=\mathrm{diag}(2,-2,-2,-2)$, the first two terms in Eq.~(\ref{dirac hamiltonian}) lead to the Dirac equation in Eq.~(\ref{DP equation}) with $\mu'=0$ and $t_3=0$. [The `standard (Dirac) representation' of the gamma matrices~\cite{BagrovGitman+2014} is obtained by the unitary transformation $U^\dagger \bm{\alpha} U $ where $U=e^{-i\tau_3\pi/4}$, which corresponds to a $\pi/2$-rotation in the $\tau$-subspace.] 
Note that $t_3\neq0$ 
is beyond the Dirac equation, as is common for gapped Dirac materials~\cite{Zhang2009}.
We note that while this term does not break the algebraic structure of the Dirac equation its presence leads to measurable consequences, e.g., in the determination of the energy of a bound state in an impurity potential (see below).

Finally, in the absence of magnetic fields  $[\gamma^\mu,\gamma^\nu]F^{\mu\nu}=4\bm{\gamma}\cdot{\bm E}$, establishing equivalence between Eqs.~(\ref{DP equation}) and~(\ref{dirac hamiltonian}) with $\mu'=\mu$. Note that the magnetic part of the Pauli term $[\gamma^\mu,\gamma^\nu]F^{\mu\nu}$ 
leads to an anomalous $\tau_3\otimes\bm{\sigma} \cdot\bm{B}$ coupling, which also enters the low-energy description of LHPs placed in the magnetic field $\bm{B}$~\cite{PhysRevB.107.125201,PhysRevLett.130.106901}. However, the precise magnitude of this magnetic term in MAPbBr$_3$ is currently  undetermined, and therefore, we leave the corresponding investigation to future studies.

In what follows, we  illustrate the physical implications of the Pauli term for LHPs. To this end,
we solve the stationary Dirac-Pauli equation  
\begin{equation}
    H\Psi=\mathcal{E}\Psi
    \label{time independant sch eqn},
\end{equation}
where $\Psi$ is the 4-component Dirac spinor, $\Psi^T=(\psi_{\mathcal{\Uparrow}\uparrow},\psi_{\mathcal{\Uparrow}\downarrow},\psi_{\mathcal{\Downarrow}\uparrow},\psi_{\mathcal{\Downarrow}\downarrow})$, and $\mathcal{E}$ is the energy. [Here, the first (last) two elements of the spinor are typically referred to as electron (hole) components.] Specifically, we solve Eq.~(\ref{time independant sch eqn}) with the Dirac-Pauli Hamiltonian for two standard physical problems: 
{\color{black} bound states and scattering. Specifically, we discuss binding of a charge carrier by an attractive potential and scattering of free charge carriers off a potential step (Klein paradox; see Fig.~\ref{fig:R vs barrier height}(a)). To guide experimental studies of the Pauli term, we further discuss spin effects in scattering and formation of excitons. The coupling between spin and the electric field is a hallmark feature of the Pauli coupling, providing a clear pathway for its detection in a laboratory.

\vspace{1em}

\noindent {\bf Bound states near static impurities in LHPs}}

\noindent In this subsection, we investigate how the bound states of an electron in the vicinity of a spherically symmetric local impurity potential, $\phi$, are modified by the Pauli term. To this end, we focus on the effective behavior of electrons by integrating out the hole degrees of freedom following the procedure of Ref.~\cite{Keldysh1964} (see Supplementary Note 1). {\color{black}This approach resembles Foldy-Wouthuysen transformation in QED~\cite{bjorken1964relativistic}, which is a perturbative study with the expansion parameter $\epsilon/\Delta$ ($\epsilon$ is the energy of the bound state). Perturbative approach to bound states is natural in LHPs due to its `defect tolerance', i.e., only shallow states are observed in the badgap~\cite{Jin2020}. 

By doing so, we frame the problem in the form of the Schr{\"o}dinger equation for electrons  that in the leading (in the powers of $1/\Delta$) order reads } 
\begin{equation}
       \bigg( -\frac{\nabla ^2}{2 m_e^*}+q \phi(r)+V_0+H_{P}\bigg)\psi(r) =\epsilon  \psi(r),
\label{Schrodinger equation for electrons}
\end{equation}
where
$\psi^T=(\psi_{\mathcal{\Uparrow}\uparrow},\psi_{\mathcal{\Uparrow}\downarrow})$ is the electronic wavefunction; $\epsilon=\mathcal{E}-\Delta/2$ is the corresponding energy; $m_e^*=(8a^2t^2/\Delta+t_3 a^2/2)^{-1}$ is the effective mass, note that the parameter $t_3$ enters only in the effective mass, otherwise its presence does not affect our results. {\color{black}The terms $V_0$ and $H_P$ originate from the coupling between the conduction and  valence bands. $V_0=-\frac{q^2 \phi(r)^2}{\Delta}+\frac{2 \epsilon}{\Delta} q \phi(r)$ is the beyond-Coulomb potential energy of an electron in the electric field generated by the impurity for $\mu=0$~\cite{Keldysh1964}}; $H_P$ is the sum of terms arising from Pauli coupling:
\begin{equation}
  H_{P}=\frac{\big|\mu\bm{E}(\bm{r})\big|^2}{\Delta}-\frac{2at\mu}{\Delta}{\bm{\nabla}}\cdot\bm{E}(\bm{r})-\frac{8at\mu}{\Delta} \frac{|\bm{E}(\bm{r})|}{r} \bm{S}\cdot\bm{L}.
\label{Pauli term in main text}
\end{equation}

The first term in $H_P$ mimics the effect of polarizability of quasi-electrons. This term is non-vanishing even if $t=0$, implying that this polarizability of electrons inside lead-halide perovskites can be traced back to the atomic polarizability of the underlying lattice ions (primarily Pb$^{2+}$), see also the Discussion section below. The $\bm{\nabla}\cdot\bm{E}(\bm{r})$ term is referred to as the Darwin term; in QED, this term is interpreted as coming from the jitter motion ($Zitterbewegung$) of a relativistic electron on the background of a non-uniform electric field~\cite{bjorken1964relativistic}. The last term in $H_P$ describes the spin-orbit coupling (SOC).

If $\mu\neq0$, SOC appears as $\sim \mu |\bm{E}(\bm{r})|\bm{S}\cdot\bm{L} /(r\sqrt{m_e^*\Delta})$ in the effective Schr{\"o}dinger equation~(\ref{Schrodinger equation for electrons}) of the Dirac-Pauli system. Hence, its influence on the states and energies is different  when compared to a Dirac system of an electron in a given electric field. For example, in a hydrogen atom SOC has the paradigmatic form $\sim q|\bm{E}(\bm{r})|\bm{S}\cdot\bm{L}/(r m_e\Delta)$~\cite{bjorken1964relativistic}. We see that the above presented terms have identical functional dependence. However, their strengths are different with the ratio $\mu\sqrt{m_e^*\Delta}$. A similar conclusion can be reached also for the Darwin-like term.

\begin{figure}
    \centering
    \includegraphics[width=\linewidth]{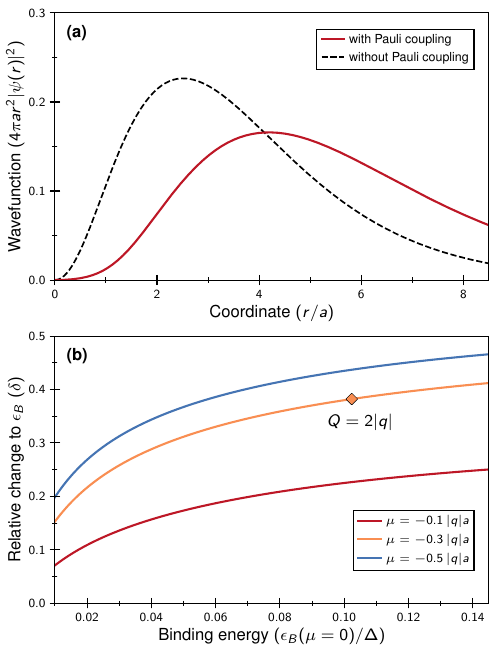}
    \caption{{\bf Bound state problem in presence of Pauli coupling.} (a) Ground state wavefunction with and without the Pauli term for $\frac{\mu}{a q}= 0.3$ {\color{black}assuming the electric field from Eq.~(\ref{eq:electric_field_impurity}) with $Q=2|q|$}. (b) Relative change in the binding energy of the electron due to the Pauli coupling  plotted as a function of the binding energy with $H_P=0$. Note that the Pauli coupling leads to a decrease in the binding energy. Different curves are for different values of the effective dipole moment, $\mu$; the curve in the middle is calculated with the parameters of MAPbBr$_3$. {\color{black}The marker corresponds to the parameters used in panel (a).}}
    \label{fig:Zitterbewegung}
\end{figure}

To illustrate the effect of $ H_{P}$ on the properties of bound states, we investigate the ground state of an electron interacting with a model Gaussian defect whose charge density is chosen as  
\begin{equation}
\rho(r)=\frac{Q}{(a \sqrt{\pi})^3} e^{-\frac{r^2}{a^2}},
\end{equation} where $Q$ is the charge of the defect.
The width of the Gaussian is determined by the lattice constant, $a$, to model a single-site defect of a semiconductor. We consider a defect with $Q>0$, which binds electrons. However, the hole-electron symmetry allows us to also easily understand what happens for $Q<0$ from our calculations.

The charge density produces a radial electric field given by
\begin{equation}
E(r)=\frac{Q}{4 \varepsilon \pi^{3/2} a r^2} \bigg( a \sqrt{\pi}  \;
 \mathrm{erf}(r/a)-2re^{-\frac{r^2}{a^2} }\bigg),
 \label{eq:electric_field_impurity}
\end{equation}
where $\varepsilon$ is the (dynamic) dielectric constant of the medium (we take $\varepsilon=5$ in our numerical calculations~\cite{Jiang2019}) and $\mathrm{erf}$ is the standard error function. Once the electric field is known, we solve the Schr{\"o}dinger equation~\eqref{Schrodinger equation for electrons} numerically by diagonalizing the corresponding matrix (see Methods). Specifically, we calculate the ground state energy of the system with and without $H_{P}$, and study the difference introduced by $H_{P}$.

The ground state of Eq.~(\ref{Schrodinger equation for electrons}) has $L=0$, hence, the spin-orbit coupling term does not play a role in Eq.~(\ref{Pauli term in main text}) and can be ignored. In other words, effects of SOC can only be seen in {\color{black}the states with $L\neq0$}. A study of the ground state allows us to demonstrate the effect of the first two terms in $H_P$: the polarizability and the Darwin terms. Note that in LHPs, the parameter $\mu$ is negative, which implies that the matrix $H_P$ is  positive-definite, and hence, that the ground state energy is increased by the presence of the Pauli coupling.

We show our findings in Fig.~\ref{fig:Zitterbewegung}. First, we plot the electronic density in  Fig.~\ref{fig:Zitterbewegung}(a). We see that the wave function becomes spatially broader as the Pauli term repels the electron. This change in the electronic density affects other observables as well, for example the binding energy $\epsilon_B>0$ ($\epsilon_B=-\epsilon$) of the electron-defect system. 
Indeed, Fig.~\ref{fig:Zitterbewegung}(b) confirms that the Pauli term leads to a noticeable  reduction of $\epsilon_B$ (see also Supplementary Figure 1). In the figure, we plot the relative decrease in binding energy:
\begin{equation}
\delta=
\frac{\epsilon_B(\mu=0)-\epsilon_B(\mu)}{\epsilon_B(\mu=0)}.
\end{equation}

 We conclude that the Pauli coupling in lead-halide perovskite can have a measurable effect on the energies of electrons in the vicinity of defects. This effect scales with the parameter $\mu$, which depends on the chemical structure of the material.
 As we expect that $\mu$ can be affected by one's choice of the halogen atom, further studies are needed to determine the lead-halide perovskite with the most pronounced effect of Pauli coupling for tabletop simulations of the corresponding physics.

 \vspace{1em}
\noindent {\color{black}
{\bf Spin structure of excitons in LHPs}

\noindent LHPs are known for their anomalously fast photon emission at cryogenic as well as at room temperatures~\cite{Rain2016,Becker2018}. It has been suggested that this is because excitons in LHPs have a unique fine structure with an optically active (“bright”) ground state. Such level ordering can be obtained by introducing a Rashba-type spin-orbit interaction~\cite{Becker2018,Sercel2019}.
In general, there is a consensus that Rashba coupling is featured prominently in LHPs, however it is usually introduced phenomenologically to account for experimental observations. The microscopic origin of Rashba coupling is a subject of debate, especially at high temperatures where LHPs have a nominally cubic structure~\cite{Etienne2016,Sajedi2020,Niesner2016}.

With the Pauli term, Rashba coupling follows naturally, with a coefficent determined by the parameters of the perovskite structure (cf.~Ref.~\cite{PhysRevLett.130.106901}):
\begin{equation}
H_{\mathrm{Rashba}}=-\frac{4 a t\mu}{\Delta}\tau_3\otimes \mathbf{E}\cdot(\mathbf{k}\times\bm\sigma),
\label{eq:HSOC}
\end{equation}
where $\mathbf{E}$ is the electric field explicitly breaking inversion symmetry. The Rashba energy that corresponds to $H_{\mathrm{Rashba}}$ is
\begin{equation} E_{\mathrm{Rashba}}\simeq\frac{8 a^2 t^2 \mu^2 \langle \mathbf{E}^2\rangle}{\Delta^2}m_e^*.
\label{eq:deltaE_Rashba}
\end{equation}
According to Ref.~\cite{Becker2018}, the lowering of the ``bright'' state is given by $\delta E_X\simeq - E_{\mathrm{Rashba}}$.
To compensate for the standard singlet-triplet splitting ($\simeq 0.3$meV~\cite{Becker2018}) due to the electron-hole exchange interaction~\cite{Pikus1971}, one should have $\delta E_X\simeq 1$meV.   
Under the hypothesis of local polar fluctuations~\cite{Yaffe2017}, this implies the fields $|\mathbf{E}|\sim 0.6$V/nm.

While internal fields of this strength are reasonable on the atomic scales \cite{Wei2023}, they are quite considerable and are yet to be experimentally confirmed.  
This requirement for the strength of these fields is high because of the factor $1/\Delta^2$ in Eq.~(\ref{eq:deltaE_Rashba}),
which follows from the use of second-order perturbation theory.
Surprisingly, it turns out that Pauli term can lower the optically active exciton state at a lower order of perturbation theory, thus producing the desired level splitting with lower electric fields ($|\mathbf{E}|\sim 0.3$V/nm). Indeed, the effective polarizability of a charge carrier due to Pauli coupling implies a van-der-Waals-like lowering of the energy 
by 
\begin{equation}
\delta E_X\simeq -\frac{\mu^2 \mathbf{E}^2}{\Delta},
\end{equation}
which enters in the first order in $1/\Delta$, in contrast to the result in Eq.~(\ref{eq:deltaE_Rashba}).

We conclude that the Pauli coupling may play a role in the observed fast photon emission in LHPs.  
Note that all derivations in this work apply to a cubic phase of LHPs. However, the polarizability of lead ions is a local property of the material, thus, we expect that our discussion applies also at lower temperatures, where LHPs crystals are tetragonal or orthorhombic.

\vspace{1em}
\noindent {\bf Klein problem in LHPs} }

\noindent Klein paradox refers to a counterintuitive behavior of scattering amplitudes off a step potential whose height is larger than twice the particle's rest mass. In the first quantization picture, the paradox manifests itself in 
the seemingly unphysical values for the reflection and transmission coefficients. In particular, the Klein's solution suggests that the 
transmitted current does not vanish even when the height of the barrier approaches infinity~\cite{Klein1929,Sauter1932} (for more details see Refs.~\cite{hansen1981klein,Dombey1999,Calogeracos1999}).

\begin{figure}
    \centering
    \includegraphics[width=\linewidth]{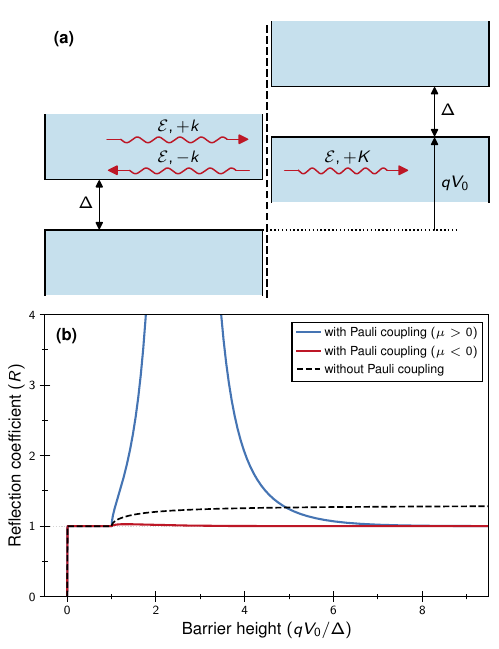}
    \caption{{\bf Klein problem in LHPs.} (a) Schematic of the Klein problem:  Electrons of energy $\mathcal{E}$ and momentum $k$ are incident on a  potential barrier of height $V_0$ {\color{black}(recall that the associated electric field is $E_x = -V_0\delta(x)$)}. If the value of $V_0$ is larger than the energy gap, $\Delta$, between the conduction and valence bands, then Klein paradox occurs -- particles with momentum $K$ appear in the high potential region. (b) Reflection coefficient as a function of the (dimensionless) barrier height with and without Pauli coupling assuming that $\mathcal{E}=\Delta/2+0.01\,\text{eV}$. The other parameters correspond to MAPbBr$_3$ with $\mu<0$. For $\mu>0$, only the sign of $\mu$ is changed in comparison to the case with $\mu<0$.}
    \label{fig:R vs barrier height}
\end{figure}

Direct experimental observation  of this physics in quantum electrodynamics would require access to extremely high electric fields not available in modern laboratories~\footnote{The exception is electron beams in aligned crystals, which however can only be used for indirect studies of Klein paradox~\cite{Nielsen2023}}.
This motivates searches for tabletop set-ups that would realize Dirac's equation, such as gapless Dirac materials~\cite{Katsnelson2006,Young2009} or photonic superlattices~\cite{Longhi2010}.
The most direct analogy to the Klein paradox in its original formulation however requires working with gapped Dirac semiconductors such as LHPs.
 The corresponding bandgap $\Delta$ then would play the role of a pair-creation energy $2mc^2$ in QED, where $m$ is the mass of the particle that is being created. As the gap $\Delta$ can be about six orders of magnitude lower than $2mc^2$ the corresponding effects in Dirac semiconductors are much more accessible as compared to QED. 
 Nevertheless, it should be noted that the very existence of an energy gap of a few eV may already result in a considerable experimental obstacle for the observation of Klein paradox in LHPs. Still, we find it fitting to start the discussion of the effect of the Pauli term by considering the corresponding solution of the Dirac equation.

For simplicity, we assume that $t_3=0$ here~\footnote{{\color{black}Otherwise, we should explicitly consider other electronic bands whose presence leads to $t_3\neq0$}}. Similar to the original formulation of the problem by Klein,
we consider the one-dimensional potential  \begin{equation}\phi(x)=V_0 \Theta(x),
\label{eq:step_potential}
\end{equation}
where $\Theta(x)$ is the Heaviside step function, and $V_0$ is the strength of the potential. For $\mu\neq0$, the associated electric field $E_x=-\partial \phi/\partial x = -V_0\delta(x)$ enters the Dirac equation via the Pauli term.  This leads to a term with a $\delta$-function in Eq.~(\ref{dirac hamiltonian}), giving rise to a discontinuity in the spinor components at $x=0$, which 
can be treated as a specific boundary condition, see Supplementary Note~2 and Ref.~\cite{calkin1987proper}. 

We solve the Dirac-Pauli equation for each spinor component and arrive at the following expressions for the reflection and transmission coefficients (see Supplementary Note 2):
\begin{equation} R=\bigg|\frac{r-1}{r+1}\bigg|^2,\qquad T=\frac{\mathrm{Im} \Lambda}{\lambda} \frac{4 |r|^2}{\big|r+1\big|^2}e^{\frac{\mu V_0}{a t}},
\label{eq:reflectivity}
\end{equation}
 where $\lambda=4i atk/(\Delta+2\mathcal{E})$ and $\Lambda=4i atK/(\Delta+2\mathcal{E}-2qV_0)$, $r=\lambda e^{-\frac{\mu V_0}{a t}}/\Lambda $ and $k$ and $K$ are the momenta of the charge carrier for $x<0$ and $x>0$. For $\mu\to0$, these expressions turn into the original Klein solution~\cite{Calogeracos1999}.  Non-vanishing values of $\mu$ lead to an exponential dependence on $V_0$ in Eq.~(\ref{eq:reflectivity}). Subject to the sign of $\mu$, the exponent $e^{-\frac{\mu V_0}{a t}}$ that appears in the parameter $r$ is either increasing or decreasing, leaving marked imprints on the transmission and reflection coefficients.
 
 When $\mu<0$ as in LHPs, the Klein paradox is modified in comparison to the case when there is no Pauli coupling. Curiously, the introduction of the Pauli term suppresses the anomalous ($R>1$) Klein reflectivity in the asymptotic limit of large barrier heights, $qV_0 \rightarrow \infty$. Figure~\ref{fig:R vs barrier height}(b)  illustrates this for the parameters of MAPbBr$_3$.  The behavior of the reflection coefficient for $\mu>0$ is even more intriguing as it features
 a resonance-like divergent behavior, which can be interpreted as a resonant enhancement of particle-antiparicle creation within the potential barrier. This might be indicative of a novel vacuum breakdown channel that has previously been overlooked (see Supplementary Note 3 for additional details). We conclude that for the either sign of $\mu$, Pauli coupling strongly affects Klein's physics.   

\vspace{1em}
 {\color{black}
 \noindent \textbf{Spin effects in scattering}
 
\noindent Without doubt, modifications to Klein's physics is of general interest, however, their observation appears challenging due to the energy gap in LHPs. As we discuss below, the effects of Pauli coupling on scattering are more readily studied through the spin degree of freedom. To illustrate this, let us assume that a sample of LHPs has an energy barrier described by the potential $\phi(x)=v_0f(x)$, where $v_0$ ($f$) is the strength (shape) of the barrier. By assumption, the function $f(x)$ has a finite support parametrized by $d$, i.e., $f(x)=0$ if $|x|>d/2$. One realization of the potential $\phi(x)$ is an n-p-n junction~\cite{Xu2022}.    
 Unlike the Klein scenario, the potential barrier here is weak, $v_0\ll\Delta$, allowing us to implement Foldy-Wouthuysen-like transformation and derive the equation for electron propagation, cf.~Eq.~(\ref{Schrodinger equation for electrons}):
 \begin{equation}
  \bigg( -\frac{\nabla ^2}{2 m_e^*}+W(x)+\frac{4at\mu v_0}{\Delta}\frac{\mathrm{d}f}{\mathrm{d}x}\sigma_zk_y\bigg)\psi(\mathbf{r}) =\epsilon  \psi(\mathbf{r}),
  \label{eq:scattering_after_Klein}
 \end{equation}
 where we chose the $z$-axis perpendicular to the scattering plane; $W(x)$ is the spin-independent part of the potential (see Supplementary Note 1); $k_y$ is the quantum number that parametrizes the momentum along the $y$-direction. 

The Pauli coupling in Eq.~(\ref{eq:scattering_after_Klein})
 turns the energy barrier into a magnetic impurity, which affects spin-up and spin-down electrons differently, allowing one to manipulate the spin of an electron. For example, given a spin-unpolarized current for $x<0$, one can generate a spin-polarized current for $x>d$. Indeed, let us consider the transmission coefficient (see, e.g., Ref.~\cite{Fad64})
 \begin{equation}
 T_{\pm}=\left|1+\frac{i m_e^*}{k_x}\int\mathrm{d}x W_{\pm}(x)e^{-ik_x x}\psi_1(x)\right|^{-2}.
 \end{equation}
  Here, $\pm$ corresponds to the spin of the incoming electron; $W_{\pm}= W\pm 4atv_0\mu f'k_y/\Delta$. The function $\psi_1$ is the solution to Eq.~(\ref{eq:scattering_after_Klein}) that satisfies the integral equation
  \begin{equation}
  \psi_1(x)=e^{ik_x x}-\frac{2m_e^*}{k_x}\int_x^\infty \mathrm{d}y\sin[k_x(x-y)]W_{\pm}(y)\psi_1(y),
  \end{equation}
  where the momentum $k_x$ parametrizes the incoming flux. As $W_{+}\neq W_-$, we conclude that $T_+\neq T_-$, which leads to the spin polarization of the outgoing current: $|T_+-T_-|/(T_++T_-)$. To measure it in LHPs, one can use the standard methodology established for other semiconductors~\cite{Belykh2019}.

  The simplest estimate for $T_\pm$ is obtained in the quantum tunneling regime using the WKB approximation, $T^{\mathrm{WKB}}_\pm\sim \exp(-2\int_{x_1}^{x_2} \sqrt{2m_e^*(W_\pm-\epsilon)})$, where $x_1$ and $x_2$ are classical turning points. This expression allows one to optimize $f(x)$ for the problem at hand. It is worth noting that $T^{\mathrm{WKB}}_+\neq T^{\mathrm{WKB}}_- $ only for potentials that break mirror symmetry, i.e., if $f(-x)\neq f(x)$. This is in line with 
the recent observation of spin polarized currents in chiral LHPs at room temperature in the absence of external magnetic fields~\cite{Kim2021}. Our results demonstrate that similar observations can be made in non-chiral LHPs in the presence of an energy barrier with broken mirror symmetry.

 }

\vspace{1em}
\noindent{\large \bf Discussion}

\noindent It was noted by Foldy that the presence of the Pauli term implies finite polarizability of the particle in question~\cite{PhysRevLett.3.105}. Indeed, for a Hamiltonian of the form $H=\beta m +\bm{\alpha}\cdot\bm{k}+i \mu \bm{\gamma} \cdot \bm{E}$, the energy of the zero-momentum mode reads 
\begin{equation}
\mathcal{E}^{k=0} = \sqrt{m^2 + \mu^2 E^2} \sim m + \frac{\mu^2 E^2}{2 m}.
\end{equation}
The energy shift quadratic in the electric field forces an interpretation in terms of the polarizability of the particle, which is not associated with any actual structural deformation. We interpret the Pauli term in LHPs in a similar manner: 
it describes an effective polarizability of an electron (and hole) quasiparticle in LHPs, which on the microscopic level can be traced back to the local atomic polarizability of  the perovskite lattice,  which is large due the presence of Pb$^{2+}$ ions. In other words, the spin-electric term $\mu\ \tau_1 \otimes\bm\sigma \cdot \bm{E}$
originates from a coupling between valence and conduction bands that is mediated by the electric field even in the limit $k=0$ due to the lattice polarizability.
The fact that the value of $\mu$ was found to be considerable in LHPs~\cite{PhysRevLett.130.106901} implies that quasi-electrons dressed by the high polarizability of lead ions obeys the Dirac-Pauli equation.

To the best of our knowledge, LHP is the only known family of materials where the Pauli term cannot be neglected, providing a convenient testing ground for exploring intriguing theoretical concepts from high-energy physics. In this paper, we  demonstrate some basic physical manifestations of Pauli coupling in the problems related to scattering off potentials and the spectrum of electronic bound states. However, there are other scenarios where the effects of the Dirac-Pauli physics could be interesting to explore.
{\color{black}For example, we note that the Pauli term naturally leads to spin dephasing in the vicinity of impurities, see spin-orbit coupling in Eq.~(\ref{Pauli term in main text}). Inclusion of the Pauli term is hence necessary for understanding  spin relaxation mechanisms in LHPs~\cite{Schwinger1948,Elliott1954,Yafet1963}.
 These are under intensive experimental and theoretical investigations at the moment~\cite{Zhou2020,Xu2024} due to potential spintronic applications of LHPs~\cite{Privitera2021}.}
Finally, recent observations of the dynamical Schwinger effect~\cite{lorenc2024dynamical} opens an opportunity to simulate analogues of  relativistic strong-field effects, and to investigate (both theoretically and experimentally) the Pauli coupling in the presence of strong laser fields.

  \vspace{3em}

  \noindent {\large \textbf{Methods}}

\noindent We study Eq.~(\ref{Schrodinger equation for electrons}) 
for a spherically symmetric potential  (i.e., \textcolor{black}{$\phi(\bm{r})=\phi(r)$ and $\bm{E}(\bm{r})=\bm{E}(r)$}), therefore, it is convenient to work in spherical coordinates \textcolor{black}{$(\bm{r}\equiv\{r,\Omega\})$} where it can be written as
\begin{equation}
    \bigg( -\frac{1}{2 m_e^*} \frac{d^2}{dr^2}+q \phi(r)+V_0+H_{P}\bigg)U(r)=\epsilon\,U(r)
\label{Schrodinger equation in polar coordinates}
\end{equation}
with $\psi(\textcolor{black}{\bm{r}}) = (4\pi)^{-1/2}U(r)/r$. To solve this equation numerically, we consider the system in a box potential,~i.e., $r\in[0,R]$ and $R\gg a$, and expand $U(r)$ over a complete basis given by this potential
\begin{equation}
    U(r)=\sqrt{\frac{2}{R}}\sum_n \alpha_n\sin(k_n r)\,.
    \label{eq_basis}
\end{equation}    
The boundary condition at $r=R$ defines the allowed values of $k_n$: $\sin(k_n R)=0\longrightarrow k_n=\pi n/R$; the prefactor in Eq.~(\ref{eq_basis}) follows from the normalization condition\textcolor{black}{: $\int_{0}^{R}\sin(k_i r)\sin(k_j r)\,\mathrm{d}r=\frac{R}{2}\delta_{i,j}$.}
Note that we have used the fact that the three-dimensional electronic wave function \textcolor{black}{$\psi(\bm{r})$} must be finite at the origin for physical potentials, which implies that $U(0)=0$.

Finally, we substitute Eq.~\eqref{eq_basis} into Eq.~\eqref{Schrodinger equation in polar coordinates}\textcolor{black}{, project the resulting expression onto the basis functions $\sin(k_j r)$, and} obtain the linear system of equations
\textcolor{black}{
\begin{equation}
    \forall\,j:\quad
    \sum\limits_{i} H_{i,j}\alpha_i=\epsilon\sum\limits_{i}\bigg[\delta_{i,j}-\frac{2}{\Delta}\expval{q\phi(r)}_{i,j}\bigg]\alpha_i\,,
    \label{eq_linear_system}
\end{equation}
where 
\begin{equation}
H_{i,j}=\bigg[\frac{k_i^2}{2 m_e^*}\delta_{i,j}+ \expval{q\phi(r)}_{i,j} - \frac{1}{\Delta}\expval{(q \phi(r))^2}_{i,j}+
    \expval{H_P}_{i,j}
 \bigg] \nonumber.
\end{equation}
We have defined the matrix elements as $\expval{f(r)}_{i,j}\equiv\frac{2}{R}\int_0^R\sin(k_j r)f(r)\sin(k_i r)\,\mathrm{d}r$.} 

To diagonalize Eq.~\eqref{eq_linear_system}, \textcolor{black}{ we} introduce a momentum cutoff $\forall\,i:k_i\textcolor{black}{\leq} k_{\text{max}}$ with $k_{\text{max}} a\gg 1$, which leads to a finite-dimensional matrix equation whose eigenvalues can be found numerically. We adjust the binding energy of the electron by changing the charge of the defect, $Q$, see Supplementary Figure 1. Then, we produce Fig.~\ref{fig:R vs barrier height}(b) by connecting $Q$ to the binding energy of the system without the Pauli term.

  \vspace{3em}
  
 \noindent \textbf{Data and code availability} The `minimum set' that is necessary to interpret, verify and extend the data presented in Figs.~\ref{fig:Zitterbewegung} and~\ref{fig:R vs barrier height} is available at \url{https://git.ista.ac.at/mmaslov/dirac_pauli_LHP}. 

 \vspace{1em}

  \noindent \textbf{Author Contributions}
Z.A. and A.G.V. conceived and designed the project. A.S.K. and A.G.V. performed the theoretical calculations and corresponding analysis under the supervision of M.L. and Z.A. A.S.K. and M.M. performed numerical calculations under the supervision of A.G.V. All authors contributed to the interpretation of the results and wrote the paper.

\vspace{1em}
\noindent The authors declare no competing interests.

\newpage 

\widetext 

\section{Supplementary Note 1}
Here, we present technical details for bound-state calculations presented in the main text. For convenience, we write
the equation $ H\Psi= \mathcal{E} \Psi$, where $H$ is a $4\times4$ matrix and $\Psi=\begin{pmatrix}\psi_{\textcolor{black}{1}}\\\psi_{\textcolor{black}{2}}\end{pmatrix}$\textcolor{black}{, where} ($\textcolor{black}{\psi_1\equiv}\psi_{\mathcal{\Uparrow}}$ \textcolor{black}{is} the electron manifold; $\textcolor{black}{\psi_2\equiv}\psi_{\mathcal{\Downarrow}}$ \textcolor{black}{is} the hole manifold) as
\begin{subequations}
\begin{equation}
\Tilde{H}_{11} \psi_{\textcolor{black}{1}}+H_{12} \psi_{\textcolor{black}{2}}=0
\label{subequation1}
\end{equation}
\begin{equation}
H_{21} \psi_{\textcolor{black}{1}}+\Tilde{H}_{22} \psi_{\textcolor{black}{2}}=0
\label{subequation2}
\end{equation}
\end{subequations}
where $\Tilde{H}_{11}=H_{11}-\mathcal{E},\Tilde{H}_{22}=H_{22}-\mathcal{E}$. Here $H_{ij}$ is the $ij^{\textcolor{black}{\text{th}}}$ block of the Hamiltonian \textcolor{black}{$(2\times2\text{ matrix})$}.
Substituting $\psi_{\textcolor{black}{2}}$ from Eq.~\eqref{subequation2} into Eq.~\eqref{subequation1} and introducing \textcolor{black}{the commutator }$[H_{\textcolor{black}{12}}\textcolor{black}{,\tilde{H}_{22}}]$, it takes the form 
\begin{equation}
\textcolor{black}{\big(}H_{12}H_{21}-\Tilde{H}_{22}\Tilde{H}_{11}-[H_{12}\textcolor{black}{,}\Tilde{H}_{22}]\Tilde{H}_{22}^{-1}H_{21}\textcolor{black}{\big)}\psi_{\textcolor{black}{1}}=0.
    \label{eqn for electron}
\end{equation}
We make this equation dimensionless by dividing it with the mass term of the Dirac equation, $\Delta$:
    \begin{equation}
        (H_0 + H')\psi_{\textcolor{black}{1}}=0, \qquad \mathrm{where} \qquad H_0=\frac{1}{\Delta}(H_{12}H_{21}-\Tilde H_{22}\Tilde{H}_{11}) \qquad \mathrm{and} \qquad  H'=-\frac{1}{\Delta}[H_{12}\textcolor{black}{,}\Tilde{H}_{22}]\Tilde{H}_{22}^{-1}H_{21}.
    \end{equation}
In our analysis, $H_0$ is the leading-order Hamiltonian \textcolor{black}{(}in the powers of $\Delta$\textcolor{black}{) and} $H'$ is  $1/\Delta^2$-perturbation (recall that $H_{\textcolor{black}{22}}^{-1} \sim 1/\Delta$ ).

Let us calculate the Hamiltonian $H_0$ for a slowly changing electric field whose effect can be taken into account within the local density approximation. To this end, we derive
\begin{equation}
    H_{12}H_{21}
=-4 a^2 t^2 \nabla^2 - 2 a t \mu \bm{\nabla}\cdot\bm{E}(\bm{r}) - i2at\mu \bm{\sigma}\cdot(\bm{\nabla}\cross\bm{E}(\bm{r}))
+i4at \mu \bm{\sigma}\cdot(\bm{E}(\bm{r})\cross\bm{\nabla})+\mu^2 \big|\bm{E}(\bm{r})\big|^2,
\label{1st term}
\end{equation}
\begin{equation}
-\Tilde{H}_{22}\Tilde{H}_{11}
=\left( \frac{\Delta^2}{4} - \mathcal{E}^2\right) - \frac{t_3 a^2 \Delta}{4} \nabla^2+ 2 \mathcal{E} q \phi(\textcolor{black}{\bm{r}})
 - (q \phi(\textcolor{black}{\bm{r}}))^2 +\frac{t_3 a^2 }{4} \left(q\phi(\textcolor{black}{\bm{r}})\nabla^2 - q\nabla^2 \phi(\textcolor{black}{\bm{r}}) \right).
\label{2nd term prototype}
\end{equation}
Note that $\bm{\nabla}\cross\bm{E}=0$ since we are dealing with a time-independent problem; further, $\bm{\sigma}\cdot(\bm{E}\cross\bm{\nabla})=i2 \frac{|\bm{E}(\textcolor{black}{\bm{r}})|}{r} \bm{S}\cdot\bm{L}$ for a spherically symmetric problem with $\bm{S}=\bm{\sigma}/2$ ($\bm{L}=-i(\mathbf{r}\cross \bm{\nabla})$) being the standard spin (angular momentum) operator. 
 Hence,  equation~\eqref{1st term} becomes 
 \begin{equation}
     H_{12}H_{21}
=-4 a^2 t^2 \nabla^2 - 2 a t \mu \bm{\nabla}\cdot\bm{E}(\bm{r}) 
+\mu^2 \big|\bm{E}(\bm{r})\big|^2 - 8at\mu \frac{|\bm{E}(\bm{r})|}{r}  \bm{S}\cdot\bm{L}.
\label{1st term main}
 \end{equation}
Furthermore, we have that
$\int\mathrm{d}\mathbf{r} f(\mathbf{r})\nabla^2 \phi(\textcolor{black}{\bm{r}})f(\mathbf{r})=\int\mathrm{d}\mathbf{r} f(\mathbf{r}) \phi(\textcolor{black}{\bm{r}})\nabla^2 f(\mathbf{r})
$ for any real square integrable function $f$. Therefore, the last term in Eq.~(\ref{2nd term prototype}) can be neglected for our purposes.

Introducing the notation: $\mathcal{E}=\frac{\Delta}{2}+\epsilon$, where $\epsilon\ll \frac{\Delta}{2}$, we write down the Hamiltonian $H_0$ as 
\begin{equation}
H_0=-\frac{4 a^2 t^2}{\Delta}\nabla^2-\frac{t_3 a^2}{4}\nabla^2+\frac{\mu^2 \big|\bm{E}(\bm{r})\big|^2}{\Delta}-\frac{2at\mu}{\Delta}\bm{\nabla}\cdot\bm{E}(\bm{r})
- \frac{8at\mu}{\Delta}\frac{|\bm{E}(\bm{r})|}{r}  \bm{S}\cdot\bm{L}-\epsilon +q \phi(\textcolor{black}{\bm{r}})-\frac{\textcolor{black}{(q\phi(\bm{r}))^2}}{\Delta}+\frac{2 \epsilon}{\Delta} q \phi(\textcolor{black}{\bm{r}}) ,
\label{zeroth order Hamiltonian}
\end{equation}
where $-\frac{2at\mu}{\Delta}\textcolor{black}{(}\bm{\nabla}\cdot\bm{E}(\bm{r})\textcolor{black}{)}$ resembles the standard Darwin term and $- \frac{8at\mu}{\Delta}\frac{|\bm{E}(\textcolor{black}{\bm{r}})|}{r}  \textcolor{black}{(}\bm{S}\cdot\bm{L}\textcolor{black}{)}$ describes spin-orbit coupling. 
 To write the corresponding Schr{\"o}dinger equation, we define the effective mass of electrons in the material, $\frac{1}{m_e^*}=\frac{8a^2t^2}{\Delta}+\frac{t_3 a^2}{2}$, so that
 \begin{equation}
    \bigg( -\frac{\nabla ^2}{2 m_e^*}+q \phi(\textcolor{black}{\bm{r}})\textcolor{black}{-\frac{(q \phi(\bm{r}))^2}{\Delta}+\frac{2 \epsilon}{\Delta} q \phi(\bm{r})}+H_{P}\bigg)\psi(\textcolor{black}{\bm{r}})=\epsilon \psi(\textcolor{black}{\bm{r}}),
\label{Schrodinger equation}
\end{equation}
where
\begin{equation}
  H_{P}=\frac{\mu^2}{\Delta}\big|\bm{E}(\bm{r})\big|^2-\frac{2ta\mu}{\Delta}[{\bm{\nabla}}\cdot\bm{E}(\bm{r})]-\frac{8at\mu}{\Delta} \frac{|\bm{E}(\bm{r})|}{r} \bm{S}\cdot\bm{L}
\label{Pauli term}
\end{equation}
is the result of the Pauli term; $\psi(\textcolor{black}{\bm{r}})$ is the electronic wavefunction.
 Using the values $a=0.586 \,\,\text{nm} $, $t=0.6\,\,\text{eV}$, $t_3 =0.9\,\,\text{eV}$ and $\Delta=2.3\,\,\text{eV}$, the effective mass of an electron is $m_e^*\simeq \textcolor{black}{0.13\,m_e}$, where $m_e$ is the rest mass of a free electron. 

\begin{figure}
    \centering
    \includegraphics[width=0.5\linewidth]{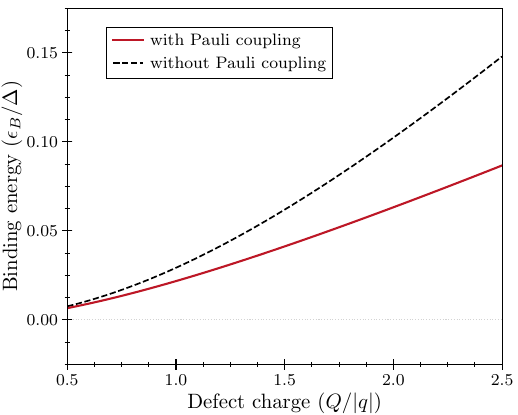}
    \caption{{\bf Binding energy of the electron-defect bound state}. The upper (lower) curve shows the energy without (with) Pauli coupling  as a function of the charge of the defect.}
    \label{fig:SM_energies}
\end{figure}

Finally, let us consider an electron moving in two dimensions ($xy$-plane), interacting with a potential barrier along the $x$ direction. In this case, Eq.~\eqref{Pauli term} can be written as
\begin{equation}
  H_{P}=\frac{\mu^2}{\Delta}E(x)^2-\frac{2ta\mu}{\Delta} \frac{d}{dx} E(x) + i\frac{4at\mu}{\Delta} \sigma_3 E(x) \frac{d}{dy}.
\label{Pauli term 2 D}
\end{equation}
Note that $\psi_{\mathcal{\uparrow}}$ and $\psi_{\mathcal{\downarrow}}$ acquire different energies due to the last term in $H_{P}$, which acts like an effective magnetic field on the pseudo-spins. The pseudo spins can undergo precession about the axis of this effective magnetic field, which may lead to spin polarization of the outgoing current (see the main text for details).

\section{Supplementary Note 2}
{\it Formalism.--} In this Supplementary note, we derive Eq. (15) of the main text. To this end, we consider a one-dimensional problem with $\mathcal{E}>\Delta/2$: 
\begin{equation}
    \left(\frac{\Delta}{2}+ V(x)-\mathcal{E}\right)\psi_{\mathcal{\Uparrow}\uparrow}+\left(\mu E_x -2 a t \frac{d}{dx}\right)\psi_{\mathcal{\Downarrow}\downarrow}=0
\label{2D Dirac 1}
\end{equation}

\begin{equation}
     \left(-\frac{\Delta}{2}+ V(x)-\mathcal{E}\right)\psi_{\mathcal{\Downarrow}\downarrow}+\left(\mu E_x +2 a t \frac{d}{dx}\right)\psi_{\mathcal{\Uparrow}\uparrow}=0
\label{2D Dirac 2}
\end{equation}
where $V(x)$ is the potential energy given by $V(x)=q V_0 \Theta(x)$. The strength of the potential is determined by $qV_0>0$ and its shape is described by the Heaviside step-function $\Theta(x) =
\left\{
	\begin{array}{ll}
		1  & \mbox{if } x \geq 0 \\
		0 & \mbox{if } x < 0
	\end{array}
\right. .$   The associated electric field is $E_x = -V_0\delta(x)$, where $\delta(x)$ is the Dirac delta function. Due to the presence of the $\delta$-function, the spinor components are no longer continuous at $x=0$. Below, we find the corresponding discontinuity condition.

 To regularize the problem for $\mu=0$, one can consider the $\delta$-function as a limit 
$\delta(x) = \lim_{w \to 0} f(x)$ where $f(x) =
\left\{
	\begin{array}{ll}
		0  & \mbox{if } |x| > w \\
		\frac{1}{2 w} & \mbox{if } |x| < w
	\end{array}
\right.$~\cite{calkin1987proper}. As we shall show, one can do the same for $\mu\neq0$. The only complication in comparison to Ref.~\cite{calkin1987proper} is the existence of $V(x)$ that should change linearly in the region $|x| < w$ to make the limiting procedure consistent. To demonstrate that this linear change does not affect the discussion in Ref.~\cite{calkin1987proper}, we integrate Eq.~\eqref{2D Dirac 1}  from $-w$ to $w$
\begin{equation}
\left(\frac{\Delta}{2}-\mathcal{E}\right) \int_{-w}^{w}\psi_{\mathcal{\Uparrow}\uparrow} \,dx +\int_{-w}^{w}\psi_{\mathcal{\Uparrow}\uparrow} \ V(x) \,dx =
\mu V_0  \int_{-w}^{w}\frac{1}{2 w}\psi_{\mathcal{\Downarrow}\downarrow} \,dx +2at\int_{-w}^{w} \frac{d \psi_{\mathcal{\Downarrow}\downarrow} }{dx} \,dx.
\label{2D Dirac 1 integral}
\end{equation}
The left-hand-side of this equation vanishes in the limit $w\to0$, also the potential $V(x)\ll f(x)$ for $|x|<w$ allowing us to approximate the wave functions $\psi_{\mathcal{\Downarrow}\downarrow}$ and $\psi_{\mathcal{\Uparrow}\uparrow}$ for $|x|<w$ with those for $V(x)=0$. With these observations in mind, one follows Ref.~\cite{calkin1987proper} to derive the boundary condition
\begin{equation}
   \frac{\psi_{\mathcal{\Uparrow}\uparrow}(w)}{\psi_{\mathcal{\Uparrow}\uparrow}(-w)}=e^{\frac{\mu V_0}{2 a t}}, \qquad  \frac{\psi_{\mathcal{\Downarrow}\downarrow}(w)}{\psi_{\mathcal{\Downarrow}\downarrow}(-w)}=e^{-\frac{\mu V_0}{2 a t}} .
  \label{boundary condition 1}  
\end{equation}

{\it Solution. --} We look for a scattering solution to Eqs.~\eqref{2D Dirac 1} and~\eqref{2D Dirac 2} in the form $\Psi^T=(\psi_{\mathcal{\Uparrow}\uparrow},\psi_{\mathcal{\Downarrow}\downarrow})$
\begin{equation}
\Psi(x<0)=A \begin{pmatrix} 1
\\ \lambda
\end{pmatrix}e^{i k x} + B \begin{pmatrix} 1
\\ -\lambda
\end{pmatrix}e^{-i k x}, \qquad \Psi(x>0)= C \begin{pmatrix} 1
\\ \Lambda
\end{pmatrix} e^{iKx},
\end{equation}
  where $k=\sqrt{\mathcal{E}^2-\frac{\Delta^2}{4}}/(2at)$ and  $K=\sqrt{(\mathcal{E}-qV_0)^2-\frac{\Delta^2}{4}}/(2at)$
  represent the electron momenta for $x<0$ and $x>0$, respectively;
  $\lambda=i 2atk/\left(\frac{\Delta}{2}+\mathcal{E}\right)$ and $\Lambda=i 2atK/\left(\frac{\Delta}{2}+\mathcal{E}-qV_0\right)$. 
   The coefficients $A$, $B$ and $C$ are derived from the discontinuity relation at $x=0$:
\begin{equation}
    C = (A-B) e^{\frac{-\mu V_0}{2 a t}} \frac{\lambda}{\Lambda}, \qquad  C = (A+B) e^{\frac{\mu V_0}{2 a t}} .
  \label{discontinuity criterion 1}  
\end{equation}

  The corresponding reflection and transmission coefficients are calculated using the current densities of incoming and outgoing waves. To define them, we first write Eqs.~\eqref{2D Dirac 1} and~\eqref{2D Dirac 2} in a compact form for $|x|\to\infty$
\begin{equation}
     \bigg[\sigma_z\frac{\Delta}{2}+\big(V(x)-i\partial_t\big)\mathbb{I}-2 i at\sigma_y \partial_x\bigg]\Psi =0.
\label{2D Dirac combined}
\end{equation}
Note that this equation does not have any information about the Pauli term, and hence we can use the standard definition of the Dirac current: 
by first multiplying Eq.~(\ref{2D Dirac combined}) by $\Psi^\dagger$ from the left and the Hermitian transpose of this equation with  $\Psi $ from the right, and then subtracting the outcomes, we derive
\begin{equation}
    \partial_t\big(\Psi^\dagger\Psi\big)+2at \partial_x\big(\Psi^\dagger \sigma_y\Psi\big)  =0 .
\label{continuity equation}
\end{equation}
This equation defines the current density as $j=2at(\Psi^\dagger \sigma_y \Psi)$. The reflection and transmission coefficients can be now defined as $R=\frac{j_\text{reflected}}{j_\text{incident}}=\left|\frac{B}{A}\right|^2$  and  $T=\frac{j_\text{transmitted}}{j_\text{incident}}=\left|\frac{C}{A}\right|^2 \frac{\mathrm{Im}{\Lambda}}{\lambda}$, respectively, leading to the result presented in the main text.

\section{Supplementary Note 3}

\begin{figure}
    \centering
    \includegraphics[width=\linewidth]{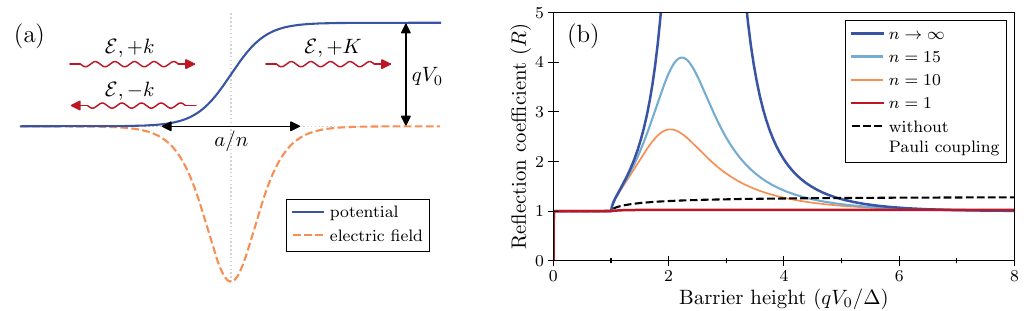}
    \caption{ {\bf Klein paradox for a smooth potential barrier.} (a) Sketch of a smooth potential barrier from Eq.~\eqref{tanh}. (b) The reflection coefficient as a function of the (dimensionless) barrier height for different steepness of the potential, which is characterized by $n$. The figure is for $\mu>0$.}
    \label{fig:smoothness}
\end{figure}

\begin{figure}
    \centering
    \includegraphics[width=\linewidth]{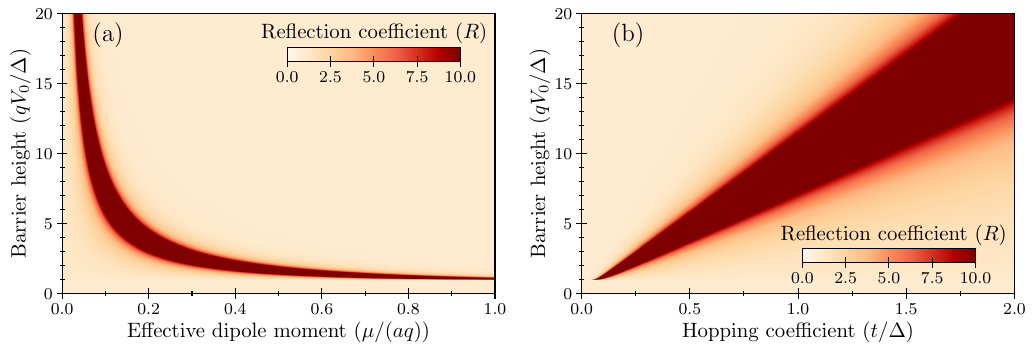}
    \caption{{\bf Reflection coefficient in Klein paradox.}  Reflection coefficient ($R$) as a function of the (dimensionless) barrier height $q V_0/\Delta$ and (a) (dimensionless) effective dipole moment $\mu/(aq)$; (b) (dimensionless) hopping coefficient $t/\Delta$.}
    \label{fig:solar figure}
\end{figure}

Here, we consider the case of $\mu>0$.  In comparison to the case of $\mu<0$ considered in the main text, the major difference is the occurrence of a resonance at the barrier height corresponding to vanishing denominators in Eq.~(15) of the main text, i.e., for $r=-1$, see~\ref{fig:smoothness}(b). It can be demonstrated that this resonance is not a mathematical artefact of the employed step potential by using in Eqs.~\eqref{2D Dirac 1} and~\eqref{2D Dirac 2} a hyperbolic tangent potential
\begin{equation}
    V(x)=\frac{q V_0}{2} \Bigg[\mathrm{tanh}\bigg(\frac{nx}{a}\bigg)+1\Bigg]
    \label{tanh},
\end{equation}
standard in the studies of Klein paradox~\cite{Sauter1932}.
Here, $n$ characterises the steepness of the potential, see~\ref{fig:smoothness}(a). Finally, we show in \ref{fig:solar figure} how the properties of the resonance are modified as a function of other parameters.

\bibliography{mybib}

\end{document}